# Subjectivity Classification using Machine Learning Techniques for Mining Feature-Opinion Pairs from Web Opinion Sources

Ahmad Kamal
Department of Mathematics
Jamia Millia Islamia (A Central University)
New Delhi – 110025, India

**Abstract**

Due to flourish of the Web 2.0, web opinion sources are rapidly emerging containing precious information useful for both customers and manufactures. Recently, feature based opinion mining techniques are gaining momentum in which customer reviews are processed automatically for mining product features and user opinions expressed over them. However, customer reviews may contain both opinionated and factual sentences. Distillations of factual contents improve mining performance by preventing noisy and irrelevant extraction. In this paper, combination of both supervised machine learning and rule-based approaches are proposed for mining feasible feature-opinion pairs from subjective review sentences. In the first phase of the proposed approach, a supervised machine learning technique is applied for classifying subjective and objective sentences from customer reviews. In the next phase, a rule based method is implemented which applies linguistic and semantic analysis of texts to mine feasible feature-opinion pairs from subjective sentences retained after the first phase. The effectiveness of the proposed methods is established through experimentation over customer reviews on different electronic products.

## 1. Introduction

With the exponential growth of World Wide Web and rapid expansion of e-commerce, web opinion sources such as merchant sites, forums, discussion groups and blogs are used as a platform by individual users to share experiences or opinions. Online merchant sites provide space for customers to write feedback about their product and services, as a result number of customer reviews grow rapidly for each product. Such reviews are useful for customers in making purchase decision regarding a product based on the experiences of the existing users, whereas on the other hand, it helps product manufacturers in assessing strength and weaknesses of their products from the perspective of end users. Such information is very useful in developing marketing and product development plans.

Recently, feature based opinion mining technique is gaining momentum in which every granule of customer reviews are processed to identify product features and user opinions expressed over them. However, customer reviews may contain both subjective and objective sentences. Subjective sentences represent user's sentiment, feeling, belief, rants, etc. In contrast, objective contents represent factual information. Consider the following review sentences:

- The battery life of this camera is very *good*.
- Camera is a g*ood* device for capturing photographs.

Both sentences contains opinion bearing word *good*, despite first sentence is subjective and second one is objective in nature. Thus, the target of subjectivity classifications is to restrict unwanted and unnecessary objective texts from further processing. However, classifying a sentence as either subjective or objective is a non-trivial task due to non availability of training dataset. Annotated sets of subjective and objective sentences are difficult to obtain and requires lots of manual processing and thus time consuming [1].

The aim of the current work is to propose methods for identifying subjective sentences from customer reviews for mining product features and user opinions at the intersection of both machine learning and rule-based approaches. In the first phase of the proposed approach, a supervised machine learning technique is applied for subjectivity or objectivity classification for each word of a review sentence, and thereafter the probability of the inscribing sentence to be subjective or objective is calculated using a unigram model. In the next phase, extracted subjective sentences are taken as input by rule-based method which applies linguistic and semantic analysis of texts to identify information components. Initially, an information components are extracted to fill a template *<f, m, o>*, where *f* represents a product feature, *o* represents an opinion expressed over *f*, and *m* is a modifier used to model the degree of expressiveness of *o*.

Since, for a product feature, different users may express same or different opinions and a single user with in a review document may express opinions on different features, the basic assumption is that a simple frequency based summarization of the extracted feature-opinion pairs is not suffice to express their reliability. In line with [2], noisy extraction of pairs is handled by calculating *reliability score* for every candidate feature-opinion pair. The value of *reliability score* determines the reliability of an opinion expressed over a product feature.

The remaining paper is structured as follows: section 2 presents a brief review of the existing work in subjectivity classification, product feature and opinion identification. Section 3 presents the architectural and functional details of the proposed system. The experimental setup and results evaluation are presented in section 4. Finally, section 5 concludes the paper.

## 2. Related Work

The purpose of subjectivity and objectivity classification in opinion mining research is to distinguish between factual and subjective remarks present in customer reviews. Such classification of texts can be performed both at document and sentence levels. The aim of document level subjectivity classification is to identify documents containing subjective texts from large collections for further processing. Due to availability of star rated (1 to 5 stars) customer reviews at merchant sites, divisions among subjective and objective documents are simple. Higher star rated document can be placed in subjective class, whereas lower star rated document can be assigned to objective class [1, 12]. However, a study in [1] revealed that many documents contain combination of both subjective and objective sentences. Subjective texts may also include some factual contents. For example, a movie review usually considered as subjective document (as it reflects sentiments and feelings of its viewers) may contain factual description regarding actors, plot, and list of theaters where the movie is currently playing. On the other hand, objective documents such as newspaper article may enclose subjective texts. Wiebe *et al* in [3] have reported 44% subjective sentences in objective news collection after discarding editorial and review articles. Thus, for better classification performance, sentence level subjective or objective analysis is proposed by various researchers. Many research efforts acknowledge the presence of *adjective* in a sentence as a good indicator for sentence subjectivity. Study in [4] revealed that adjectives are statistically, significantly and positively correlated with subjective sentences in the corpus on the basis of the log-likelihood ratio test. If there exist at least one adjective in the sentence, the probability of a sentence being subjective is 55.8%, despite of more objective than subjective sentences in the corpus. In [5], Hatzivassiloglou & Wiebe study the effects of dynamic adjectives, semantically oriented adjectives, and gradable adjectives on a simple subjectivity classifier and establish that they are strong predictors of subjectivity. Their prediction method for subjectivity states that a sentence is classified as subjective if at least one member of a set of adjective occurs in the sentence otherwise objective. In addition to use *adjective* as a subjectivity indicator in a sentence, study in [6] reported that *noun* can also be used for subjectivity determination. Their experiment using naïve Bayes algorithm achieved a precision of 81% for sentence level subjectivity classification task.

Apart from analyzing subjective or objective sentences in customer reviews, another important task in the field of feature based opinion mining research is to identify product features and user's opinions expressed on them. In [7], Hu and Liu have used an unsupervised method and applied a three-step process for features and opinions extraction. In [8], semi-supervised technique double propagation is proposed to extract features and opinion words using seed opinion lexicon. Further, extracted features and opinions are exploited for identifying new features and opinions. In [9], an unsupervised mutual reinforcement approach is proposed which clusters product features and opinion words simultaneously by fusing both content and link information. Clustering can be very useful for domains where the same feature word is referred by its various synonym. Authors in [10], proposed a supervised approach for movie reviews that apply grammatical rules to identify feature-opinion word pairs. Since, a complete opinion along with its relevant feature is always expressed in one sentence [11], the feature and opinion pair extraction can be performed at sentence level to avoid their false associations.

## 3. Proposed Subjectivity Classification and Feature-Opinion Pair Mining Method

In this section, architecture and functional details of the proposed subjectivity or objectivity classification and feature-opinion pair mining methods are presented. Fig. 1 presents the complete architecture of the proposed system, which consists of five different functional components – *data crawler, document pre-processor, subjectivity/objectivity analyzer, feature and opinion learner,* and *feasibility analyzer*. Further details about these components are presented in the following subsections.

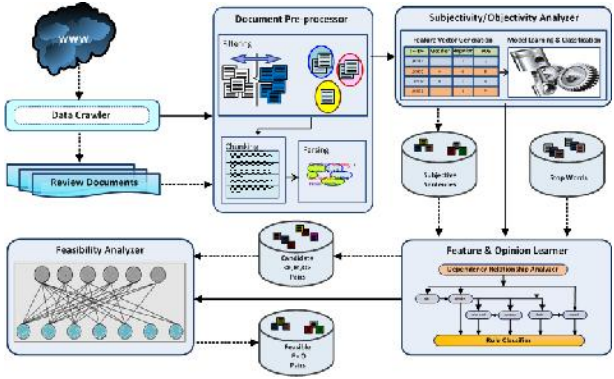

**Fig. 1:** Architecture of the proposed system

### 3.1 Data Crawler and Document Pre-processor

For a target review site, the data crawler retrieves review documents and store locally after filtering markup language tags. The filtered review documents are divided into manageable record-size chunks whose boundaries are decided heuristically based on the presence of special characters. For facilitating subjectivity classification and information component extraction, linguistic as well as semantic analysis of text is performed by assigning Parts-of-Speech (POS) tags to every word of a review sentence using POS analyzer. The POS tag reflects the syntactic category of the word and plays vital role in identification of relevant features, opinions and modifiers from review sentences. In this proposed work, POS based filtering mechanism is applied to avoid unwanted texts from further processing.

### 3.2 Subjectivity/Objectivity Analyzer

Machine learning approaches are likely to provide more accurate classification results, and very useful in learning patterns for identification of subjectivity/objectivity in customer reviews. A supervised machine learning technique is proposed for subjectivity or objectivity classification of each word present in a review sentence, and thereafter the probability of the enclosing sentence to be either subjective or objective is calculated using a unigram model. However, for machine learning application to work effectively, the important task is to engineer set of features and their formulation in a way to produce best classification result. For this purpose, feature vector generator is implemented and attributes such as *term frequency*, *parts of speech*, o*pinion indicator seed word*, *position*, *negation, and modifier* are used to build a binary classification model for characterization of candidate subjective and objective unigrams from a review sentence. In line with [12], the formulations of different features are presented in the following section.

**TF-IDF:** It combines the frequency of a unigram in a particular review document with its occurrence in the whole corpus. It is calculated using equation (1), where *f* is the frequency count of the unigram in the review document, *s* is the size of the review document in terms of words, $N_f$ is the number of review documents in the corpus containing the unigram, *C* is the total number of review documents in the corpus.

$$TF - IDF = \frac{f}{s} \times (-\log_2 \frac{N_f}{c}) \qquad (1)$$

**Position:** It determines the position of the occurrence of candidate unigram in a review sentence. Sometime, position of the unigram plays an important role in deciding sentence subjectivity. The position attribute is set to -1, 0, and 1 in case candidate unigram occurs in the beginning, in-between, and end respectively of the enclosing review sentence.

**POS:** Part-of-Speech (POS) information is one of the most promising among all features, and used commonly in subjectivity detection. A large number of researches reveal *adjectives* as a good indicator of opinion. Further, nouns (*problem*, *pain*, *issue*, *love*), verbs (*degrade, like*) and adjectives (*hard*, *pretty*) can also be used for subjectivity determination of a word. Feature value is set to *A*, *D*, *N*, *V*, and *E* in case candidate unigram *is adjective*, *adverb*, *noun*, *verb,* and *any other* respectively.

**Opinion Indicator Seed Word:** Opinion indicator seed words are commonly used by reviewers for expressing positive or negative sentiment regarding product features or services and can be used as a good indicator for subjectivity determination. For example, set of positive seed words {*amazing*, *awesome*, *beautiful*, *decent*, *nice*, *excellent*, *good}* and set of negative seed words {*bad*, *bulky*, *expensive*, *faulty*, *horrible*, *poor*, *stupid}.* The feature value is set to 1 in case candidate unigram belongs to positive seed set, and set to 0 if it belongs to negative seed set.

**Negation:** Presence of negation is also treated as an important clue for subjectivity detection. In case the candidate unigram is a negation word, the feature value is set to 1 otherwise set to 0.

**Presence of Modifier:** A modifier word usually adverb is used to express the degree of expressiveness of opinion in review sentences. If the candidate unigram is found to be a modifier, then the feature attribute is set to 1 otherwise set to 0.

**Class Attribute:** This attribute is defined only for the training set of unigrams. If a unigram is subjective, then its value is set to '*S*', otherwise '*O*'.

Table 1 shows a partial list of the candidate subjective and objective unigrams. Table 2 shows an exemplar feature vector to represent candidate subjective/objective unigram of a review sentence.

**Table 1:** Exemplar subjective and objective unigrams

| Subjective unigram | Objective unigram |
|---|---|
| amazing, beautiful, cheap, decent, effective, fantastic, good, happy, impress, jittery, light, madly, nice, outstanding, perfect, quick, responsive, sharp, terrible, ultimate, wonderful. | access, because, chance, default, entire, few, go, half, inside, job, keep, know, last, matter, new, only, past, quality, read, several, text, use, version, was, young. |

**Table 2:** Feature vectors for subjective/objective unigrams

| TF-IDF | Position | POS | Opinion indicator seed word | Negation | Modifier |
|---|---|---|---|---|---|
| 0.0058 | 1 | A | 1 | 0 | 0 |
| 0.0110 | 0 | D | 0 | 1 | 0 |
| 0.0232 | 1 | N | 0 | 0 | 0 |
| 0.0067 | 0 | D | 0 | 0 | 1 |
| 0.0044 | 1 | E | 0 | 0 | 0 |
| 0.0412 | 0 | A | 1 | 0 | 0 |
| 0.0032 | 0 | D | 0 | 1 | 0 |
| 0.0352 | -1 | N | 0 | 0 | 0 |
| 0.0033 | 0 | D | 0 | 0 | 1 |
| 0.0062 | 0 | A | 0 | 0 | 0 |

The proposed method for subjectivity and objectivity determination works in two phases – *model learning* and *classification*. The first phase, also called training phase, uses feature vectors generated from training dataset to learn a classification model, which is later used to identify subjective unigrams in new dataset. The second phase is centered on classification of subjective unigrams from test dataset using the learned model. To determine the subjectivity of a review sentence, it is tokenized into unigrams and the class of each token is determined using the trained model. Finally, the sentence is considered as a subjective sentence if the predicted class for any of token is subjective. For implementation of the classification model, the naïve Bayes algorithm implemented in WEKA [13] is used due to its best performance.

### 3.3 Feature and Opinion Learner

This module is implemented as a rule-based system, and accepts subjective POS tagged review sentences as input along with dependency relationships information between words. To tackle the peculiarity and complexity of review documents, various rules are defined to access different sentence structures for identification of information components embedded within them. Table 3 represents exemplar review sentences and corresponding dependency relationships information generated by the Stanford parser [14] are shown in table 4.

**Table 3:** Example review sentences with features, opinions & modifiers

| Example Sentence | Feature | Modifier | Opinion |
|---|---|---|---|
| Samsung S5830 has a powerful battery. | Samsung S5830, battery | - | powerful |
| The picture quality is really nice, amazing and awesome. | picture quality | really | nice, amazing, awesome |

**Table 4:** Example sentences with dependency relationships

| Dependency relationships between words |
|---|
| *nn*(S5830-2, Samsung-1) *nsubj*(has-3, S5830-2) *det*(battery-6, a-4) *amod*(battery-6, powerful-5) *dobj*(has-3, battery-6). |
| *det*(quality-3, The-1) *nn*(quality-3, picture-2) *nsubj*(nice-6, quality-3) *aux*(nice-6, is-4) *advmod*(nice-6, really-5) *and*(nice-6, amazing-8) *and*(nice-6, awesome-10). |

As observed in [8], existing features can also be used to identify new feature words. In the first sentence mentioned above, the word *S5830* of the product *Samsung S5830* is the nominal subject of the verb *has* and the word *battery* is the direct object of it. Thus, *battery* can be identified as a new feature word. Further, "AMOD" relationship can be used to identify *powerful* as an opinion word. In the second sentence, the bigram *picture quality* is a product feature and can be identified using "NN" tag, whereas the word q*uality* is related to an adjective *nice* with "NSUBJ". Thus, *nice* can be identified as an opinion. Further, multiple opinion words *amazing* and *awesome* present in it can be extracted using one or more occurrence of *and* relationship with the opinion word *nice*. Here, "NN" is a noun compound modifier and "NSUBJ" is a dependency relation used in the Stanford parser. Based on these observations, various rules are designed and reported in [2, 15]. Some sample rules are presented below to highlight the function of the system.

**Rule-1:** In a dependency relation *R*, If there exist relationships *nn($w_1$, $w_2$)* and *nsubj($w_3$, $w_1$)* such that POS($w_1$)= POS($w_2$)= NN*, POS($w_3$)=VB* and $w_1$, $w_2$ is not a stop-words, Else-If, there exist a relationship *nsubj($w_3$, $w_4$)* such that POS($w_3$)=VB*, POS($w_4$)=NN* and $w_4$ is not a stop-words, then we search for *dobj($w_3$, $w_5$)* relation. If *dobj* relationship exists such that POS($w_5$)=NN* and $w_5$ is not a stop-words then ($w_1$, $w_2$) or $w_4$ as well as $w_5$ are assumed as features. Thereafter, the relationship *amod($w_5$, $w_6$)* is searched. In case of presence of *amod* relationship, such that POS($w_6$)=JJ* and $w_6$ is not a stop-words, then $w_6$ is assumed as an opinion.

**Rule-2:** In a dependency relation $R$, If there exist relationships $nn(w_1, w_2)$ and $nsubj(w_3, w_1)$ such that POS($w_1$)= POS($w_2$)= NN*, POS($w_3$)=JJ* and $w_1, w_2$ is not a stop-words. Else-If, there exist a relationship $nsubj(w_3, w_4)$ such that POS($w_3$)=JJ*, POS($w_4$)=NN* and $w_3, w_4$ is not a stop-words, then either ($w_1, w_2$) or $w_4$ is assumed as the feature and $w_3$ as an opinion respectively. Further, one or more occurrence of $and(w_3, w_k)^+$ is searched where $k \le 5$. In case of presence of *and* relationship, such that POS($w_k$)=JJ* and $w_k$ is not a stop word, then $w_k$'s are identified as opinions.

### 3.4 Feasibility Analyzer

During the information component extraction phase, various irrelevant nouns, verbs and adjectives are extracted. Sometimes, it is observed that verbs are considered as noun due to parsing error. In line with [2], noisy extractions are handled by calculating reliability score, $r_{ij}$, for every candidate feature-opinion pair ($f_i$, $o_j$), and normalizing this score using *min-max* normalization to scale it in [0, 1] as shown in equation 2, where $HS^n_{(pij)}$ denotes hub score of $p_{ij}$ after $n^{th}$ iteration (after convergence) and *NewMax* and *NewMin* values are set to 1 and 0 respectively. This metric determines the reliability of an opinion expressed over a product feature. Further details about $HS^n_{(pij)}$ can be found in [2].

$$r_{ij} = \frac{HS^n_{(p_{ij})} - \min_{xy}\{HS^n_{(p_{xy})}\}}{\max_{xy}\{HS^n_{(p_{xy})}\} - \min_{xy}\{HS^n_{(p_{xy})}\}} \times (NewMax - NewMin) + NewMin \quad (2)$$

## 4. Experimental Setup and Evaluation Results

In this section, experimental setup and evaluation results of the proposed system is presented. The data samples used in the experimental work consist of 400 review documents on different electronic product crawled from *www.amazon.com*. The dataset is crawled using *crawler4j API*[1] which is then pre-processed by some filtering to smooth the noise and chunking to decompose the text into individual meaningful chunks or sentences. Using *Stanford Parser API*[2] the text chunks are further broken down to separate the different parts of speech (POS). Standard information retrieval performance measures *precision*, *recall*, and *f-score* are used to evaluate the proposed methods, and defined in equations (3), (4) and (5) respectively.

$$precision = \frac{TP}{TP + FP} \quad (3)$$

$$recall = \frac{TP}{TP + FN} \quad (4)$$

$$f-score = \frac{2 * precision * recall}{precision + recall} \quad (5)$$

### 4.1 Evaluating Subjectivity/Objectivity Analyzer

A Java based feature vector generators is implemented to generate attributes value for each unigram present in various sentences of the data sample. A total number of 30,000 and 3,800 unigrams are generated from the training and testing datasets, respectively. A binary classification models is made consisting of two classes *subjective* and *objective*. For every unigram generated from a subjective document, the class attribute value is set to *S* otherwise it is set to *O*. From classification results, true positive *TP* (number of correct subjective/objective unigrams the system identifies as correct), false positive *FP* (number of incorrect subjective/objective unigrams the system falsely identifies as correct), and false negatives *FN* (number of correct subjective/objective unigrams the system fails to identify as correct) are obtained. These parameters are used to calculate the value of *precision*, *recall*, and *f-score* using equations (3), (4), and (5) respectively. Further, weighted average *precision*, *recall* and *f-score* values are obtained by considering weight of the two classes used for classification purpose. Weighted average value determines the relative importance of each of the *S* and *O* class on the average result.

#### 4.1.1 Analysis with Feature Attributes

This section is used to discuss the performance of most discriminative feature in the classification task. Table 5 lists the information gain ranking of various features on the basis of WEKA[3] attribute evaluator. POS information is ranked highest i.e. best discriminative features among all for subjectivity/objectivity classification followed by TF-IDF in the experiment. Fig. 2 visualizes the subjective/ objective classification of unigrams on the basis of POS information and TF-IDF using WEKA's visualizer. Majority of the *adjectives* are classified as subjective followed by *adverb*, *noun* and *verb*. Unigrams classified as subjective are represented by blue colour, and objective unigrams are visible using red colour in fig. 2.

---

[1] http://cod.google.com/p/crawler4j
[2] http://nlp.stanford.edu/software/lex-parser.shtml
[3] http://www.cs.waikato.ac.nz/ml/weka/

Table 5: Information gain ranking of features

| Features | Information Gain |
|---|---|
| POS | 0.10364911 |
| TF-IDF | 0.02714459 |
| Negation | 0.02082773 |
| Seed | 0.00113212 |
| Position | 0.00017621 |
| Modifier | 0.00000528 |

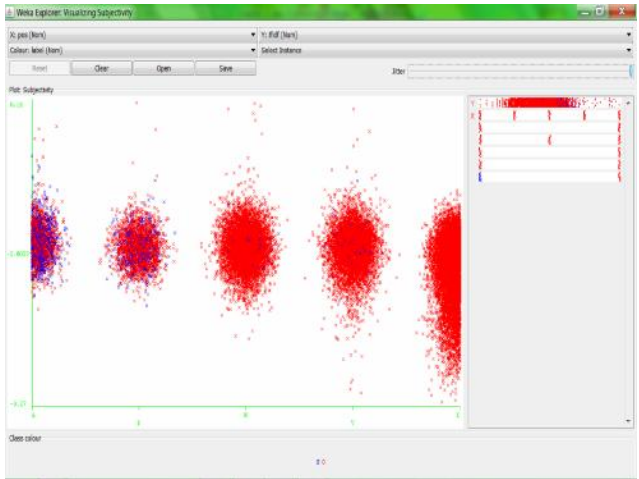

**Fig. 2:** Visualization of subjective/objective classification of unigrams based on POS information and TF-IDF values

### 4.1.2 Analysis with Classifiers

Some prominent classifiers are used for experimental purpose best suited for the classification task. Four different classifiers are considered, naive Bayes (a simple probabilistic classifier based on Bayes theorem), J48 (a decision tree based classifier), multilayer perceptron – MLP (a feed forward artificial neural network model with one input layer, one output layer and one or more hidden layers), and Bagging (a bootstrap ensemble method) and 10-fold cross-validation is used for evaluation. For determining real-time applicability of the approach using these classifiers, time consumption by the classifiers is an important concern. Table 6 shows their time consumptions during the experiment. Naive Bayes, being the simplest of all consumes 0.27 seconds, the shortest time duration of all to train the model, whereas MLP takes the longest time of 149.13 seconds. The major demerit of MLP remains in its longer training as well as testing time requirement. Since training needs to be done only once for building the model, longer training time is not a big issue, rather accuracy of the classification task is a major concern.

Table 6: Comparison of time requirements

| Classifier | Training Time (in second) | Testing Time (in second) |
|---|---|---|
| NB | 0.27 | 0.03 |
| J48 | 1.12 | 0.03 |
| MLP | 149.13 | 0.06 |
| Bagging | 3.45 | 0.05 |

### 4.1.3 Performance on Training Dataset

As discussed earlier, the training data sample used in the experiment consists of 30,000 unigrams. Fig. 3 shows the summary of correctly and incorrectly instances classified during training by various classifiers used in the experiment.

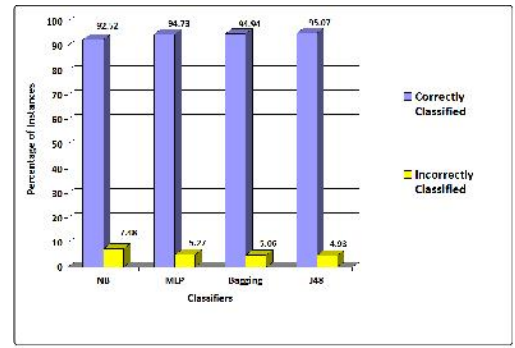

**Fig. 3**: Classification summary during training

The best classification performance is observed using J48 with correctly classified instances are 95.07%. Naïve Bayes has shown poor performance on training dataset, and percentage of correctly classified instances remains 92.52% only. As discussed earlier, standard information retrieval performance measures are used to evaluate results. In deciding the overall performance of classifiers used in the experiment, *precision*, *recall*, and *f-score* values are obtained for each of the two *subjective* and *objective* classes. As shown in table 7, for subjective class best *precision* (0.722) is obtained using J48, whereas best *recall* (0.668) and *f-score* (0.484) values are emerged from naïve Bayes algorithm. Similarly, for the *objective* class best *precision* (0.981), *recall* (1.0), and *f-score* (0.975) values are retrieved using naïve Bayes, MLP, and J48 respectively.

Table 7: Classifier's performance using IR metrics on training dataset

| Classifier | Subjective Class | | | Objective Class | | |
|---|---|---|---|---|---|---|
| | Prec. | Recall | F-score | Prec. | Recall | F-score |
| NB | 0.380 | **0.668** | **0.484** | **0.981** | 0.939 | 0.960 |
| J48 | **0.722** | 0.104 | 0.182 | 0.952 | 0.998 | **0.975** |
| MLP | 0.500 | 0.004 | 0.009 | 0.948 | **1.000** | 0.973 |
| Bagging | 0.547 | 0.229 | 0.323 | 0.958 | 0.989 | 0.974 |

Fig. 4 presents ROC curves of all four classifiers, visualizing their comparative accuracy in terms of true positive and false positive rates.

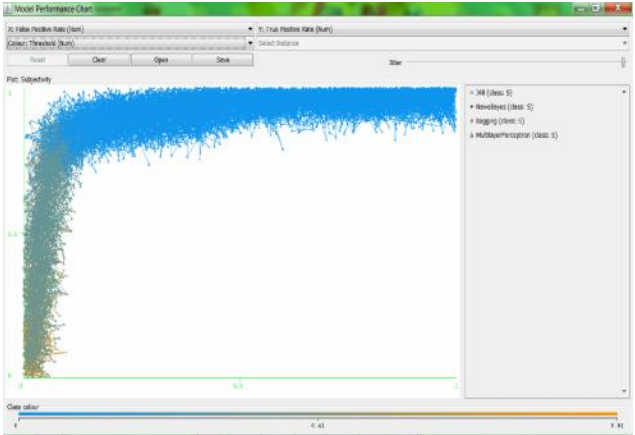

**Fig. 4**: ROC curves of classifiers for subjectivity/objectivity analysis

#### 4.1.4 Performance on Testing Dataset

Once the training phase of the proposed approach is over, trained model is used to identify subjective or objective unigrams from test dataset. 3,800 instances for testing purpose are framed. Fig. 5 shows the number of instances correctly and incorrectly classified by various classifiers on test dataset. Highest percentage of correctly classified instances i.e. 91.6% is recorded using naïve Bayes followed by J48 with 91.31% accuracy.

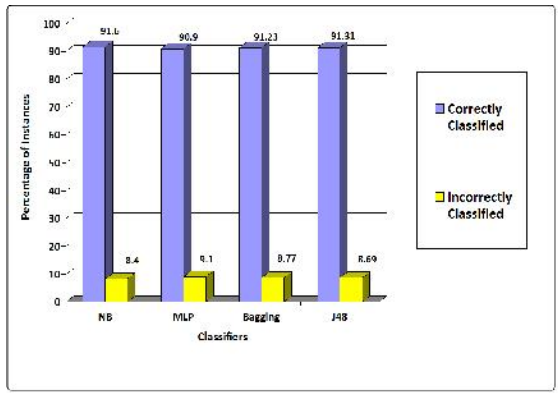

**Fig. 5:** Classification summary during testing

Similar as training, standard information retrieval performance measures are used to evaluate result on test dataset. For each of the two *subjective* and *objective* classes, *precision*, *recall* and *f-score* values are shown in table 8. For *subjective* class, best *precision* (0.789) is obtained using J48. In symmetry with training results, best *recall* and *f-score* are observed using naïve Bayes with values (0.730) and (0.610) respectively. Thus, for subjective classification, naïve Bayes reflects similar performance with labeled and unlabelled dataset used in the experiment. However, for the *objective* class best *precision* (0.972) and *recall* (0.999) is obtained using naïve Bayes and J48 respectively. However, best *f-score* (0.954) is noted using both J48 and Bagging algorithms.

**Table 8:** Classifier's performance using IR metrics on testing dataset

| Classifier | Subjective Class | | | Objective Class | | |
|---|---|---|---|---|---|---|
| | Prec. | Recall | F-score | Prec. | Recall | F-score |
| NB | 0.523 | **0.730** | **0.610** | **0.972** | 0.934 | 0.953 |
| J48 | **0.789** | 0.044 | 0.083 | 0.914 | **0.999** | **0.954** |
| MLP | 0.353 | 0.018 | 0.034 | 0.911 | 0.997 | 0.952 |
| Bagging | 0.554 | 0.012 | 0.198 | 0.919 | 0.990 | **0.954** |

In order to obtain information regarding best classifier in hand, weighted average *precision*, *recall*, and *f-score* values are obtained, in which weight (number of instances belonging to a particular class against total number of instances used for the classification purpose) of both *subjective* and *objective* classes are considered. For training dataset, it can be observed from table 9 that the best weighted average *precision* (0.949) and *recall* (0.951) values are obtained using naïve Bayes and J48 respectively. Although, highest weighted average *f-score* (0.939) is achieved using Bagging. However, the performance of naïve Bayes for weighted average *f-score* (0.935) is also comparable due to the next highest average *f-score* value. It is important to note that the better performance of the Bagging method over naïve Bayes is at the cost of the requirement of much more training time.

**Table 9:** Weighted average values

| Classifier | Weighted Average Result (over training dataset) | | | Weighted Average Result (over testing dataset) | | |
|---|---|---|---|---|---|---|
| | Prec. | Recall | F-score | Prec. | Recall | F-score |
| NB | **0.949** | 0.925 | 0.935 | **0.932** | **0.916** | **0.922** |
| J48 | 0.940 | **0.951** | 0.933 | 0.903 | 0.913 | 0.876 |
| MLP | 0.924 | 0.947 | 0.922 | 0.861 | 0.909 | 0.870 |
| Bagging | 0.937 | 0.949 | **0.939** | 0.887 | 0.912 | 0.886 |

On testing dataset, best weighted average *precision* (0.932), *recall* (0.916), and *f-score* (0.922) values are maintained by naïve Bayes. Thus, naïve Bayes has emerged as the most suitable classifier in the experiment.

### 4.2 Evaluating Feature and Opinion Learner

To the best of the knowledge, no benchmark data is available in which features and opinions are marked for electronic products. Therefore, manual evaluation is performed to monitor the overall performance of the proposed system. From the corpus of 400 review documents, a total of 45 documents (Digital Camera: 15, Laptop: 15 and Cell Phone: 15) are randomly selected consisting of 642 sentences for testing purpose. Rule-based method discussed in the section 3.3 is applied to extract feature-opinion pairs. Table 10 presents a partial list of feasible features along with opinions and modifiers.

Table 10: A partial list of extracted features, opinions and modifiers

| Product | Feature | Modifier | Opinion |
|---|---|---|---|
| Digital Camera | picture | too, very | glorious, great, excellent, fantastic |
| | view | really | bad, poor, excellent |
| | lens | too, quite | good, great, fine |
| Laptop | sound | pretty, really | great, good, perfect, clear, thin |
| | weight | extremely | light, noticeable |
| | price | very, too | higher, great, good, fantastic, reasonable |
| Cell Phone | player | enough, very | good, nice, great |
| | screen | pretty, barely, fairly, very | solid, visible, responsive, receptive |
| | software | rather | easy, slow, flimsy |

Initially, the total count obtained for *true positive* (*TP*), *false positive* (*FP*), and *false negative* (*FN*) are 251, 322, and 168 respectively. It has been observed that, direct and strong relationship between words causes extraction of various nouns (or, verbs) and adjectives that are not relevant feature-opinion pairs. As a result, counts for *FP* increase which has an adverse effect on the value of precision. To overcome this problem, a Java based *feasibility analyzer* is implemented to remove noisy feature-opinion pairs. After elimination of noisy pairs, the total count of *FP* reduces to 60. In parallel, manual collection of feature-opinion pairs from test documents are performed. Thereafter, comparing the two sets of pairs *TP*, *FP* and *FN* are calculated. Macro-averaged performance is obtained to present a synthetic measure of performance by simply averaging the result. Table 11 summarizes the performance measure values for the proposed rule-based method in the form of a misclassification matrix. The obtained *recall* (0.599) value is lower than *precision* (0.807), is an indication of system inability to extract certain feature-opinion pairs correctly.

Table 11: Performance evaluation of feature-opinion pairs extraction

| Product Category | TP | FP | FN | Precision | Recall | F-Score |
|---|---|---|---|---|---|---|
| Digital Camera | 104 | 23 | 68 | 0.818 | 0.604 | 0.694 |
| Laptop | 70 | 24 | 41 | 0.744 | 0.630 | 0.682 |
| Cell Phone | 77 | 13 | 59 | 0.855 | 0.566 | 0.681 |
| Macro-Average | 251 | 60 | 168 | **0.807** | **0.599** | **0.687** |

### 4.3 Evaluating Feasibility Analyzer

In line with [2], a Java based feasibility analyzer is implemented which compute reliability score for extracted feature-opinion pairs using equation (2). In the beginning of this step, initial score for each feature-opinion pair and review document is set to 1 and the final scores are obtained as soon as convergence of the iterative steps is reached. The convergence is reached when the score computed at two successive iterations for any review document or feature-opinion pair falls below a given threshold i.e. 0.0001. It has been observed that, most of the irrelevant noisy feature-opinion pairs lost their initial hub score, and their final score after convergence reach to a very low value tending towards zero. Table 12 represents a partial list of randomly selected noisy feature-opinion pairs discarded due to very low hub and reliability scores. Table 13 presents hub and reliability scores for some randomly selected feature-opinion pairs from different electronic products. The highest reliability score for pairs *camera-great*, *megapixel-standard*, and *phone-thin* indicates *great*, *standard* and *thin* as the most prominent qualities opined by the reviewers.

Table 12: Noisy feature-opinion pairs with low hub and reliability score values

| Feature | Opinion | Initial HS | Final HS (After Convergence) | Reliability Score (r) |
|---|---|---|---|---|
| screen refreshes | Slow | 1.00 | 0.00 | 0.00 |
| video recording | bonus | 1.00 | 0.00 | 0.00 |
| Mcafee | preinstalled | 1.00 | 0.01 | 0.00 |
| processors | Dual | 1.00 | 0.01 | 0.00 |
| speedlite | Older | 1.00 | 0.01 | 0.00 |

Table 13: Exemplar feature-opinion pairs with hub and reliability scores

| Product | Feature | Opinion | Initial HS | Final HS (After Convergence) | Reliability Score (r) |
|---|---|---|---|---|---|
| Digital Camera | camera | great | 1.00 | 18.11 | 1.00 |
| | photo | good | 1.00 | 7.76 | 0.43 |
| | picture | beautiful | 1.00 | 7.16 | 0.39 |
| | lens | great | 1.00 | 6.30 | 0.35 |
| | video | good | 1.00 | 5.78 | 0.32 |
| Laptop | megapixel | standard | 1.00 | 10.59 | 1.00 |
| | OS | great | 1.00 | 8.48 | 0.80 |
| | screen | wonderful | 1.00 | 3.43 | 0.32 |
| | keyboard | great | 1.00 | 3.27 | 0.31 |
| | price | issue | 1.00 | 2.82 | 0.27 |
| Cell Phone | phone | thin | 1.00 | 5.70 | 1.00 |
| | OS | tricky | 1.00 | 2.25 | 0.39 |
| | screen | large | 1.00 | 1.96 | 0.34 |
| | camera | good | 1.00 | 1.42 | 0.25 |
| | keyboard | awesome | 1.00 | 1.07 | 0.19 |

In last, table 14 contains top-5 authority scores with their normalized values assigned to review documents of various electronic products.

**Table 14:** Top-5 authority scored review documents

| Product | Authority Name | Initial AS | Final AS (After Convergence) | Normalized AS |
|---|---|---|---|---|
| Digital Camera | 1Canon.txt | 1.00 | 105.13 | 1.00 |
| Digital Camera | 11Kodak.txt | 1.00 | 99.33 | 0.90 |
| Digital Camera | 9Nikon.txt | 1.00 | 96.85 | 0.86 |
| Digital Camera | 21Canon.txt | 1.00 | 95.23 | 0.83 |
| Digital Camera | 13Kodak.txt | 1.00 | 94.93 | 0.83 |
| Laptop | 15Accer.txt | 1.00 | 35.78 | 1.00 |
| Laptop | 4Lenovo.txt | 1.00 | 31.60 | 0.86 |
| Laptop | 9HpReview.txt | 1.00 | 30.32 | 0.82 |
| Laptop | 83Apple.txt | 1.00 | 30.26 | 0.82 |
| Laptop | 2Apple.txt | 1.00 | 27.84 | 0.74 |
| Cell Phone | 21LGVuCUC.txt | 1.00 | 27.62 | 1.00 |
| Cell Phone | 7ATTTPhone.txt | 1.00 | 19.85 | 0.69 |
| Cell Phone | 9BlackBerry.txt | 1.00 | 19.63 | 0.68 |
| Cell Phone | 5NokiaNSma.txt | 1.00 | 18.94 | 0.66 |
| Cell Phone | 12LGVuCUC.txt | 1.00 | 18.12 | 0.62 |

## 5. Conclusions

In this paper, the design of a subjectivity/objectivity analysis system is presented based on supervised machine learning approach to identify subjective sentences in review documents. A Java based crawler is implemented to identify opinionated texts and store them locally into record-sized chunks after performing various pre-process steps. Each review sentence is tokenized into unigrams. A set of linguistic and statistical features is identified to represent unigrams as feature vectors and to learn classification models. Various classification models are considered for experimentation to establish the efficacy of the identified features for subjectivity determination. Further, a feature based opinion mining system is presented which implements a rule-based model to identify candidate feature-opinion pairs from subjective review sentences. For every extracted candidate feature-opinion pair feasibility analysis is performed by generating reliability score with respect to the underlying corpus. Standard information retrieval performance measures, including *precision*, *recall*, and *f-score* values are used to measure the accuracy of proposed methods.

## References

**Author** Ahmad Kamal is teaching Computer Science papers at the Department of Mathematics, Jamia Millia Islamia (A Central University), New Delhi, India. His qualification includes B.Sc. (H) Computer Applications,

Post Graduate Diploma in Bioinformatics (PGDBIN), and Master of Computer Applications (MCA). Currently he is pursuing Ph.D. in Computer Science from the Department of Computer Science, Jamia Millia Islamia, New Delhi. He has more than 6 years of teaching experience at undergraduate and postgraduate levels. His research interests span over the area of *opinion mining*, *natural language processing* and *machine learning*.